\documentclass{iaus}
\usepackage{graphicx}
\usepackage{amssymb}
\usepackage{amsmath}
\usepackage{epsfig}
\usepackage{bm}
\expandafter\ifx\csname package@font\endcsname\relax\else
\expandafter\expandafter
 \expandafter\usepackage
 \expandafter\expandafter
 \expandafter{\csname package@font\endcsname}%
\fi
\newcommand{\be}{\begin{equation}}
\newcommand{\ee}{\end{equation}}
\newcommand{\ba}{\begin{eqnarray}}
\newcommand{\ea}{\end{eqnarray}}

\newcommand{\prd}{{\it Phys. Rev. D}}
\newcommand{\apj}{{\it Astrophys. J.}}
\newcommand{\aj}{{\it Astron. J. (USA)}}
\newcommand{\cqg}{{\it Class. Quant. Grav.}}

\newcommand{\pla}{{\it Phys. Lett. A}}

\newcommand{\apjl}{{\it Astrophys. J. Lett.}}

\title[Gravitational bending of light by planetary multipoles]{Gravitational bending of light by planetary multipoles and its measurement with microarcsecond astronomical interferometers}
\author[Sergei Kopeikin \& Valeri Makarov]{Sergei Kopeikin$^1$ and Valeri Makarov$^2$}
\affiliation{$^1$Department of Physics \& Astronomy, University of
Missouri-Columbia, Columbia, MO 65211\\ email: {\tt kopeikins@missouri.edu} \\[\affilskip]
$^2$Michelson Science Center, California Technology Institute, Pasadena, CA 91125 \\email: {\tt vvm@caltech.edu} }
\pubyear{2008}
\volume{248}  
\pagerange{???--???}
\date{?? and in revised form ??}
\jname{A Giant Step: from Milli- to Micro-arcsecond Astrometry}
\editors{W.-J. Jin, I. Platais \& M. Perryman, eds.}

\begin{document}\maketitle
\begin{abstract}\noindent
General relativistic deflection of light by mass, dipole, and quadrupole moments of gravitational field of a moving massive planet in the Solar system is derived in the approximation of the linearized Einstein equations. All terms of order 1 $\mu$as are taken into account, parametrized, and classified in accordance with their physical origin. We discuss the observational
capabilities of the near-future optical and radio interferometers for detecting the Doppler modulation of the radial deflection, and the dipolar and quadrupolar light-ray bendings by Jupiter and the Saturn. 
\end{abstract}
Attaining the level of a microarcsecond ($\mu$as) positional accuracy and better will completely revolutionize fundamental astrometry by merging it with relativistic gravitational physics. Beyond the microarcsecond threshold, one will be able to observe a new range of celestial physical phenomena caused by gravitational waves from the early universe and various localized astronomical sources, space-time topological defects, moving gravitational lenses, time variability of gravitational fields of super-massive binary black holes located in quasars, and many others (\cite{ksge,wk,kgwinn,kopmak}). Furthermore, this will allow us to test general theory of relativity in the Solar system in a dynamic regime, that is when the velocity- and acceleration-dependent components of gravitational field (the metric tensor) of the Sun and planets bring about observable relativistic effects in the light deflection, time delay and frequency to an unparalleled degree of precision (\cite{kapjl,fk,kspv}). 

Preliminary calculations (\cite{bkk}) reveal that the major planets of the Solar system are sufficiently massive to pull photons by their gravitational fields, which have significant multipolar structures, in contrast with the Sun whose quadrupole moment is only $J_{2\odot}\leq 2.3\times 10^{-7}$ (\cite{pit}). Moreover, in the case of a photon propagating near the planet the interaction between the gravitational field and the photon can no longer be considered static, because the planet moves around the Sun as the photon traverses through the Solar system(\cite{ks,km}). The optical interferometer designed for the space astrometric mission SIM (\cite{sim}) is capable of observing optical sources fairly close in the sky projection to planetary limbs with a microarcsecond accuracy. Similar resolution can be achieved for radio sources with the Square Kilometer Array (SKA) (\cite{ska}) if it is included to the inter-continental baseline network of VLBI stations (\cite{freid}). The Gaia (\cite{gaia}) and JASMINE (\cite{jasmine}) astrometric projects represent another alternative path to microarcsecond astrometry.
It is a challenge for the SIM and SKA interferometers as well as for Gaia and JASMINE to measure the gravitational bending of light caused by various planetary multipoles and the orbital motion of the planets. This measurement, if successful, will be a cornerstone step in further deployment of theoretical principles of general relativity to fundamental astrometry and navigation at a new, exciting technological level.  

The first detection of gravitational bending of light by Jupiter was conducted in 1988 (\cite{germanteam,tl}), and the deflection term associated with the monopole field of Jupiter was determined to an accuracy of $\simeq 15\%$ to be in agreement with Einstein's general relativity theory. Later on, the Hubble Space Telescope was used to measure the gravitational deflection of light of the bright star HD 148898 as it passed within a few seconds of arc near Jupiter's limb on 24 September 1995 (\cite{hst}). Kopeikin (2001) proposed to use Jupiter's orbital motion in order to measure the retardation effect in the time of propagation of the dynamic part of gravitational force of Jupiter to photon, that appears as a small excess to the Shapiro time delay and should be interpreted as a gravimagnetic dragging of light ray caused by the orbital motion of Jupiter (\cite{kf1,kf2}). This proposal was executed experimentally in 2002 September 8, and the gravimagnetic dragging of light was measured to $\simeq 20\%$  accuracy (\cite{fk}) thus, complementing the LAGEOS measurement of the gravimagnetic field induced by intrinsic rotation of the Earth (\cite{ciuf}).

\cite{cm}  proposed to measure the deflection-of-light term associated with the axisymmetric (quadrupolar) part of Jupiter's gravitational field. Detection and precise measurement of the quadrupolar deflection of light in the Solar system is important for providing an independent experimental support for detection of dark matter via gravitational lensing by clusters of galaxies (\cite{fa}). The work by \cite{cm} can be extended in several directions (\cite{kom}). First, it assumes that light propagates in the field of a static planet while Jupiter moves on its orbit as light traverses the Solar system toward the observer. Second, \cite{cm} implicitly assumed that the center of mass of the planet deflecting light rays coincides precisely with the origin of the inertial coordinate system in the sky used for interpretation of the apparent displacements from the gravity-unperturbed (catalogue) positions of stars. This makes the dipole moment, $I^i$, of the gravitational field of Jupiter vanish, which significantly simplifies the theoretical calculation of light bending. However, the assumption of $I^i=0$ is not practical because the instantaneous position of the planet's center of mass on its orbit is known with some error due to the finite precision of the Jovian ephemeris limited to a few hundred kilometers (\cite{pit}). The ephemeris error will unavoidably bring about a non-zero dipole moment that must be included in the multipolar expansion of the gravitational field of the planet along with its mass and the quadrupole moment. 
 
The dipolar anisotropy in the light-ray deflection pattern is a spurious, coordinate-dependent effect and, hence, should be properly evaluated and suppressed as much as possible by fitting the origin of the coordinate system used for data analysis to the center of mass of the planet. Until the effect of the  gravitational dipole is properly removed from observations it will forge a model-dependent quadrupolar deflection of light because of the translational change in the planetary moments of inertia -- the effect known as the parallel-axis theorem (\cite{arn}). This translation-induced quadrupolar distortion of the light-ray deflection pattern should be clearly discerned from that caused by the physical quadrupole moment of the planet $J_2$. 

Perhaps, the SIM,
which is a Michelson-type interferometer with articulating siderostat mirrors, holds the best prospects
for precision tests of general relativity in the Solar system through gravitational bending effects. In these experimental-gravity applications
the advantages of the SIM facility are as follows:
\begin{enumerate}
\item SIM is a pointing mission. 
\item In the differential regime of 
operation the SIM interferometer is expected to achieve the unprecedented accuracy of 1 $\mu$as in a single observation on
stars separated in the sky within $\sim$2 degrees. 
\item the baseline of SIM can be rotated through 90 degrees for a dedicated observation. Two-dimensional
observations on a given set of stars are crucial for unambiguous disentangling the dipole and quadrupole deflection patterns.
\item SIM will self-calibrate its $15^\circ$-wide field of regard. This dramatically reduces
the correlated and/or systematic errors.
\item SIM can observe stars and quasars as close as several arcseconds from the planetary limb.
\end{enumerate}

Extensive discussion of various fascinating science drivers and of the evolving technical possibilities has led to a concept for the Square Kilometer Array (SKA) and a set of design goals (\cite{ska}). The SKA will be an interferometric array of individual antenna stations, synthesizing an aperture with a diameter of up to several 1000 kilometers. A number of configurations will distribute the 1 million square meters of collecting area. These include 150 stations each with the collecting area of a 90 m telescope and 30 stations each with the collecting area equivalent to a 200 meters diameter telescope. The sensitivity and versatility of SKA can provide $\sim$ 1 $\mu$as astrometric precision and high quality milliarcsec-resolution images
by simultaneously detecting calibrator sources near the target source if an appreciable component of SKA is contained in elements which are more than 1000 km from the core SKA (\cite{freid}). 

Measurement of the light bending by a moving planet with microarcsecond accuracy requires a continuous phase-referencing observation of the target and the calibrating radio sources (\cite{fk}). The main limitation of the accuracy is the tropospheric refraction which affects radio observations.  The
large-scale tropospheric refraction can be estimated by observing many
radio sources over the sky in a short period of time.  At
present the determination of the global troposphere properties can
only be estimated in about one hour, and smaller angular-scale
variations can not be determined in most cases. However, the SKA, by using observations in ten sub-arrays, on strong radio sources around the sky, will determine the tropospheric properties on time-scales which may be as short at five minutes. 

Quasars as astrometric calibrators have one peculiar property: they are variable.  The massive outflows
and shocks in the jet change the intensity and the structure
of the radio emission.  Hence, the position of the quasar reference point is variable by about 0.05 mas in most quasars.  Thus,
the calibrators used to determine the SKA astrometric precision to
better than $10~\mu$sec have jitter which is somewhat larger.  In order to reach the intended angular precision, the change in position of the calibrators must be
determined.  

In addition to various special and general relativistic effects in the time
of propagation of electromagnetic waves from the quasar to the SKA-VLBI antenna
network, we must account for the effects produced by the planetary
magnetosphere. The magnetospheric deflection estimate reveals that a single frequency observation of the light deflection will be affected by the magnetosphere at the level exceeding 1 $\mu$as. This assumes that we should observe at two widely spaced frequencies to determine and eliminate the magnetospheric effects. The noise due to turbulence in the magnetosphere (and the Earth ionosphere) may also be a limit. However, this rapidly fluctuation model is
fairly pessimistic and unlikely, and would probably average out to the
steady state model. 

Further particular details of the theoretical study of the deflection of light by quadrupole and higher gravitational multipoles can be found in papers (\cite{kkp,kom}) and references therein.

\end{document}